\documentclass[11pt]{article}
\usepackage{latexsym}
\usepackage{amssymb}
\usepackage{a4wide}
\begin{document}
\vspace{-1cm}
\begin{flushright}
IFT-UAM/CSIC-00-6\\
\end{flushright}
\vspace{1cm}
\begin{center}
{\huge\bf
K\"ahler Forms and Cosmological Solutions in Type II Supergravities
}\\[1cm]
{\Large {\bf Natxo Alonso-Alberca} and {\bf Patrick Meessen}}\\[.2cm]
{\sl Instituto de F\'{\i}sica Te\'orica\\
Universidad Aut\'onoma de Madrid, C-XVI\\
Cantoblanco, 28049 Madrid, Spain}\\[.5cm]
{\bf Abstract}
\end{center}
\begin{quote}
We consider cosmological solutions to type II supergravity theories
where the spacetime is split into a FRW universe and a K\"ahler space,
which may be taken to be Calabi-Yau.
The various 2-forms present in the theories are taken to be proportional 
to the K\"ahler form associated to the K\"ahler space.
\end{quote}
When considering cosmological compactifications
in stringy supergravity theories, one usually considers the field strength
or the field itself,
of some RR- or NS-form to be equivalent to the volume form
on some compact manifold \cite{art:Freund,art:LOW,art:LWC}.
As is well-known, on K\"ahler manifolds \cite{art:kaehler}
one can define a covariantly
constant 2-form, whose components with respect to some coordinate base
we denote by $\mathcal{J}_{mn}$, which acts like the root of a volume
form defined on the K\"ahler manifold.
Due to this volume-form-like behaviour
one can use any 2-form present in a $d$-dimensional theory
to trigger the cosmological compactification down to $d-d_{k}$, where 
$d_{k}$ is the, real, dimension of the K\"ahler manifold (Analogous
ideas have been used in spontaneous compactifications, see {\em e.g.}
\cite{art:PvN,art:romans}).
Needless to say that when $d_{k}=2$ the K\"ahler form is equivalent 
to the volume form, so that particular solutions of \cite{art:LOW}
ought to be recovered.
\par
The idea of using the K\"ahler form instead of a volume form
will be illustrated by discussing some simple cosmological 
solutions to type II supergravities. A more detailed study will
be presented elsewhere. 
Spacetime will be split into
a $d_{k}$-dimensional K\"ahler manifold
and a $(d-d_{k})$-dimensional Friedman-Robertson-Walker manifold.
In section (\ref{sec:H}) we will consider the string common sector,
where the Kalb-Ramond field will be taken to be proportional
to $\mathcal{J}$, whereas in section (\ref{sec:RR2})
we will consider the type IIA RR 2-form field strength
to be proportional to $\mathcal{J}$.
Note that analogous solutions for the type IIB RR 2-form field can be obtained
by applying S-duality to the common sector solution \cite{art:BHO}.
\par
In the string frame, the equations of motion are, where $H=dB$ ($F_{(2)}$)
is the field strength for the Kalb-Ramond
(type IIA RR 1-form {\em resp.}) field,
\begin{eqnarray}
0&=& R_{\mu\nu}
     \,-\, 2\nabla_{\mu}\nabla_{\nu}\phi
     \,+\, \textstyle{\frac{1}{4}}
               {H_{\mu}}^{\kappa\rho}H_{\nu\kappa\rho} 
     \,-\, \textstyle{\frac{1}{2}}e^{2\phi}
             \left[
                 F_{(2)\mu\kappa}{F_{(2)\nu}}^{\kappa}
                 \,-\,
                 \textstyle{\frac{1}{4}}g_{\mu\nu}F_{(2)}^{2}
             \right] \; ,\\
& & \nonumber \\
0 &=& R
      \,+\, 4\left(\partial\phi\right)^{2}
      \,-\, 4 \nabla^{2}\phi
      \,+\, \textstyle{\frac{1}{2\cdot 3!}}H^{2} \; , \\
& & \nonumber \\
0 &=& \nabla_{\mu}\left( e^{-2\phi}H^{\mu\kappa\rho}\right) 
  \;=\; \nabla_{\mu}F_{(2)}^{\mu\nu} \; ,
\label{eq:EOMstring}
\end{eqnarray}
and the Bianchi identities are $dH=dF_{(2)}=0$.
\section{Kalb-Ramond case}
\label{sec:H}
\par
The Ansatz for the metric is taken to be
\begin{equation}
ds^{2}\;=\; N^{2}(\sigma )d\sigma^{2}
      \,-\,\eta^{2}(\sigma )\overline{g}_{ij}dx^{i}dx^{j}
      \,-\, R^{2}(\sigma )h_{mn}dy^{m}dy^{n} \; ,
\label{eq:AnsatzG}
\end{equation}
where $\overline{g}_{ij}$ is a metric of constant curvature, {\em i.e.}
$R(\overline{g})_{ij}\equiv \lambda \overline{g}_{ij}$, and $h_{mn}$ is
a Ricci flat metric\footnote{Most of the equations in what follows can
easily be generalized to include a curvature for $h_{mn}$.}
on the K\"ahler manifold.
Furthermore, the dimension
of the K\"ahler manifold is $d_{k}$ and the dimension of the constant 
curvature part is $D\equiv d-1-d_{k}$, where $d$ is the dimension 
of spacetime, which will be left arbitrary.
Note that it is natural to take the internal space to be compact,
which means that it actually is a Calabi-Yau space. 
For the sake of argument though, we will always refer to the internal
space as a K\"ahler space, since this is the only property that is 
really needed.
\par
Our Ansatz for the Kalb-Ramond field, in form notation, reads
\begin{equation}
  B \;=\; f(\sigma )\, \mathcal{J} 
    \;=\; \textstyle{\frac{1}{2}} f(\sigma )
               \mathcal{J}_{mn} dy^{m}\wedge dy^{n}
    \; ,
\end{equation}
where $\mathcal{J}$ is the integrable 
almost complex structure, the K\"ahler form, on the
K\"ahler manifold and as such satisfies \cite{art:kaehler}
\begin{equation}
{\mathcal{J}^{m}}_{p}{\mathcal{J}^{p}}_{n} \;=\; -{\delta^{m}}_{n} 
   \hspace{.5cm},\hspace{.5cm}
\overline{\nabla}_{m}\, \mathcal{J}_{np}\;=\; 0 \; ,
\label{eq:KDef}
\end{equation}
where $\overline{\nabla}$ is the connection on the K\"ahler manifold.
Finally, the dilaton is taken to depend on time, $\sigma $, only.
\par
Calculating the equation of motion for the Kalb-Ramond field, one finds
that
\begin{equation}
  \dot{f} \;=\; \aleph\, Ne^{2\phi}\eta^{-D}R^{4-d_{k}} \; ,
\end{equation}
where $\aleph$ is an arbitrary constant.
Defining the field
\begin{equation}
\psi \;=\; \phi 
           \,+\, \textstyle{\frac{1}{2}}\log\left( N\right)
           \,-\, \textstyle{\frac{D}{2}}\log\left( \eta\right)
           \,-\, \textstyle{\frac{d_{k}}{2}}\log\left( R\right) \; ,
\end{equation}
introducing $M=e^{-2\psi}N$ and changing variables by $e^{2\psi}d\sigma =dt$,
one can write the equations of motion as
\begin{eqnarray}
0 &=&  \left(\log R\right)^{\prime\prime}
       \,+\, \textstyle{\frac{\aleph^{2}}{2}}R^{4} \label{eq:KR-R}\; ,\\
0 &=&  \left(\log \eta\right)^{\prime\prime}
       \,-\, \lambda\, \eta^{-2}M^{2} \; , \\
0 &=&  \left(\log M\right)^{\prime\prime}
       \,-\, D\lambda\, \eta^{-2}M^{2} \; ,  \\
0 &=& \left[ \left(\log M\right)^{\prime}\right]^{2}
      \,-\, D \left[ \left(\log \eta\right)^{\prime}\right]^{2}
      \,-\, d_{k}\left[ \left(\log R\right)^{\prime}\right]^{2}
      \,-\, \textstyle{\frac{d_{k}\aleph^{2}}{4}}R^{4} 
      \label{eq:KR-constraint}\; ,
\end{eqnarray}
where the {\em prime} indicates derivation with respect to $t$.
\par
The above equations can easily be solved to give $\lambda =0$,
$M=e^{\alpha t}$, $\eta =e^{\beta t}$ and 
\begin{equation}
 R(t)\;=\; R_{0}\, \cosh^{-1/2}\left( \aleph R_{0}^{2}t\right) \; ,
\end{equation}
which after substitution in Eq. (\ref{eq:KR-constraint}) leads 
to
\begin{equation}
\alpha^{2}\,-\, D\beta^{2} \;=\; \frac{d_{k}\aleph^{2}R_{0}^{4}}{4} \; .
\end{equation}
In these coordinates the dilaton, or equivalently the string coupling squared,
is $g_{s}^{2}= e^{2\phi}=M^{-1}\eta^{D}R^{d_{k}}$ and the Kalb-Ramond
field strength is $H=\aleph R^{4}dt\wedge \mathcal{J}$.
\par
Although there is a world of possibilities in the above
class of solutions, perhaps the most interesting case is
the one where $\beta =0$, since then the uncompactified part of
spacetime is just Minkowski space:
Taking $\alpha =1$ for convenience, one finds that the solution 
in, string, cosmological time, $\tau$, reads 
\begin{eqnarray}
ds^{2} &=& d\tau^{2}\,-\, d\vec{x}_{(D)}
       \,-\, 2R^{2}_{0} B(\tau )^{-1} \,h_{mn}dy^{m}dy^{n} \; , \nonumber \\
e^{2\phi} &=& \left( \sqrt{2}R_{0} \right)^{d_{k}} \tau^{-1}B(\tau )^{-d_{k}/2}
       \; ,\nonumber \\
H &=& 8R_{0}^{2} d_{k}^{-1/2} \tau^{-1}B(\tau )^{-2}
      \, d\tau\wedge\mathcal{J}  \; , \\
B(\tau ) &=& \tau^{2/\sqrt{d_{k}}}\,+\, \tau^{-2/\sqrt{d_{k}}} \; .
\end{eqnarray}
As one can see, this is a completely regular solution,
modulo the usual gravitational 
singularities, which smoothly interpolates between two Kasner-like 
regions \cite{art:LOW}.
From the lower dimensional point of view, the Ansatz considered above
corresponds to a solution of dilaton-gravity coupled to
moduli \cite{art:witten}, where the breathing mode, $R$, and $f$ are
the scalar fields parameterizing an $SL(2,\mathbb{R})/U(1)$ coset model.
When $d_{k}=6$, the above solutions can be obtained from the solutions
given in \cite{art:CLW} by applying an $SL(2,\mathbb{R})$ transformation
on the moduli.
\section{RR case}
\label{sec:RR2}
In much the same way as in the foregoing subsection, we can use the
RR two form in type IIA, to trigger compactification. In this case
the equations of motion and the Bianchi identity imply that 
\begin{equation}
 F_{(2)} \;=\; \aleph \mathcal{J} 
         \;=\; \textstyle{\frac{1}{2}} \aleph
               \mathcal{J}_{mn} dy^{m}\wedge dy^{n} \; .
\label{eq:F2Ansatz}
\end{equation}
Applying the same steps as in the foregoing paragraph, one finds
\begin{eqnarray}
0 &=&  \left(\log R\right)^{\prime\prime}
       \,+\, \textstyle{\frac{(d_{k}-4)\aleph^{2}}{8}}
             M\eta^{D}R^{d_{k}-4} \; ,\\
0 &=&  \left(\log \eta\right)^{\prime\prime}
       \,+\, \textstyle{\frac{d_{k}\aleph^{2}}{8}}
             M\eta^{D}R^{d_{k}-4}
       \,-\, \lambda\, \eta^{-2}M^{2} \; , \\
0 &=&  \left(\log M\right)^{\prime\prime}
       \,-\, \textstyle{\frac{d_{k}\aleph^{2}}{8}}
             M\eta^{D}R^{d_{k}-4}
       \,-\, D\lambda\, \eta^{-2}M^{2} \; ,  \\
0 &=& \left[ \left(\log M\right)^{\prime}\right]^{2}
      \,-\, D \left[ \left(\log \eta\right)^{\prime}\right]^{2}
      \,-\, d_{k}\left[ \left(\log R\right)^{\prime}\right]^{2}
      \,-\, \textstyle{\frac{d_{k}\aleph^{2}}{4}}M\eta^{D}R^{d_{k}-4}
      \,-\, D\lambda M^{2}\eta^{-2}
      \; ,
\label{eq:F2eqs}
\end{eqnarray}
\par
Looking at the above expressions, one sees that they simplify enormously
when one considers the case $d_{k}=4$:
In that case the K\"ahler breathing mode
decouples completely and one has $R=R_{0}e^{\alpha t}$. Equating also the 
powers of $M$ and $\eta$ in the equations, {\em i.e.} putting $M=\eta^{D+2}$,
one necessarily has to impose
\begin{equation}
\lambda \;=\; \frac{\aleph^{2}}{4}\left( D+3 \right) \; .
\end{equation}
The two remaining equations are implied by 
\begin{equation}
\left(\log \eta\right)^{\prime} \;=\; \pm
           \left[
               \frac{\aleph^{2}}{4} \eta^{2(D+1)}\,+\, 
                \frac{4\alpha^{2}}{D(D+3)+4}
           \right]^{1/2} \; .
\end{equation}
Now, when $\alpha\neq 0$ the solution to the above equation is complex, but
when $\alpha =0$ one finds that 
\begin{equation}
 \eta \;=\; \left(
               A \,\mp\, \frac{(D+1)\aleph^{2}}{2}t
            \right)^{-\frac{1}{D+1}} \; ,
\end{equation}
where the range of $t$ has to be chosen such that the function is 
well defined.
\par
For a stringy cosmological observer, {\em i.e.} introducing a time 
coordinate $Mdt=d\tau$, the above solution is 
\begin{eqnarray}
ds^{2} &=& d\tau^{2}
       \,-\, \left( \frac{\aleph^{2}}{2}\tau\right)^{2}d\Omega^{2}_{\lambda}
       \,-\, R_{0}^{2}\,h_{mn}dy^{m}dy^{n} \; ,\\
e^{\phi} &=& 4R_{0}^{2}\aleph^{-2}\, \tau^{-1} \; .
\end{eqnarray}
So, we see that the solution describes a 6-dimensional open 
FRW with a dilaton such that 
the string coupling strength goes to zero when $\tau\rightarrow\infty$.
\par
In order to find a solution for general $d_{k}$, we shall follow the 
same stratagem as above,
{\it i.e.} we put $R =\eta^{\alpha}$, $M=\eta^{\beta}$ and will equate the 
powers on the righthand sides of Eqs. (\ref{eq:F2eqs}). It then follows that 
$\beta = D+2+\alpha (d_{k}-4)$, using which one can calculate
\begin{eqnarray}
\lambda &=& \frac{\aleph^{2}}{16}\left(
                 d_{k}(D+3)\,+\, (d_{k}-4)^{2}
            \right) \; , \label{eq:F2lambda} \\
&& \nonumber \\
\alpha &=& -2\frac{d_{k}-4}{d_{k}(D+1)+(d_{k}-4)^{2}} \; ,\label{eq:F2alpha} \\
&& \nonumber \\
\beta &=& \frac{d_{k}(D+1)(D+2)+D(d_{k}-4)^{2}}{d_{k}(D+1)+(d_{k}-4)^{2}} \; ,
\label{eq:F2beta}
\end{eqnarray}
The dilaton is then fixed to $e^{\phi}=\eta^{2\alpha -1}$.
\par
As before, the two remaining equations are implied by
\begin{equation}
\left(\log \eta\right)^{\prime} \;=\; \pm B\, \eta^{\beta -1} \; ,
\end{equation}
where 
\begin{equation}
B^{2} \;=\; \frac{\aleph^{2}}{16}\,
            \frac{\left[ 
                     d_{k}(D+1) +(d_{k}-4)^{2} 
                  \right]^{2}}{d_{k}(D+1)+(D-1)(d_{k}-4)^{2}} \; .
\end{equation}
This then means that 
\begin{equation}
\eta \;=\; \left[ A\mp (\beta -1)B\, t\right]^{\frac{1}{\beta -1}} \; ,
\end{equation}
where $A$ is some integration constant and the range of $t$ has to be chosen
such that the above function is well-defined.
\par
Choosing the minus-sign in the last equation in order to switch to the 
cosmological time, $\tau$, one finds that $\eta = B\tau$ and the solution
reads
\begin{eqnarray}
ds^{2} &=& d\tau^{2}
       \,-\, B^{2}\tau^{2}d\Omega^{2}_{\lambda}
       \,-\, \left( B\tau\right)^{2\alpha}h_{mn}dy^{m}dy^{n} \; ,\\
e^{\phi} &=& \left( B\tau\right)^{2\alpha -1} \; .
\end{eqnarray}
Seeing Eq. (\ref{eq:F2alpha}), one may wonder whether the breathing mode
for the K\"ahler mode could grow faster than $\tau$, and thus spoil
cosmological compactification. This can however happen for $d_{k}=2$
only, in which case $\alpha =2d^{-1}$ which, seeing that we at least must
have $d=3$ in order to apply the Ansatz, is always smaller than 1.
A similar analysis for the dilaton shows that it can blow up at 
large times only when $d_{k}=2$, $d=3$
and is regular in the rest of the cases.
\section{Conclusions}
Although more general solutions are bound to exist,
the simple examples 
discussed in this work show that using a K\"ahler form instead of
a volume form is a viable option when looking for stringy cosmological
solutions.
\par
It would be interesting to see whether this kind of compactifications
survive when coupling to matter is enabled and whether
the dimension of the K\"ahler manifold will be of any importance.
It would also be interesting to generalize the above work to Heterotic
strings/M-theory (See {\em e.g.} \cite{art:Maeda})
and to investigate cosmological
solutions when other type II fields are present. 
Work in these directions is in progress.
\section*{Acknowledgments}
The authors would like to thank E. \'Alvarez, T. Ort\'{\i}n
and M.A. V\'azquez-Mozo for
useful discussions and the referee for very useful remarks.
PM would like to thank Iberdrola and the Universidad
Aut\'onoma de Madrid for their support.

\end{document}